\documentclass[pra,twocolumn,superscriptaddress,showpacs,preprintnumbers,amsmath,amssymb]{revtex4}

\usepackage{graphicx}
\usepackage{dcolumn}
\usepackage{bm}
\usepackage{enumerate}

\newcommand{\re}{\rm Re}

\begin{document}
\bibliographystyle{apsrev} 

\title{Magnetic field induced dynamical instabilities in an anti-ferromagnetic spin-1 Bose-Einstein condensate}

\author{Zhengguo Pu}
\affiliation{School of Physics and Technology, Wuhan University, Wuhan, Hubei 430072, China}

\author{Jun Zhang}
\affiliation{School of Physics and Technology, Wuhan University, Wuhan, Hubei 430072, China}

\author{Su Yi}
\affiliation{Institute of Theoretical Physics, Chinese Academy of Sciences, P.O. Box 2735, Beijing 100190, China}

\author{Dajun Wang}
\affiliation{Department of Physics, The Chinese University of Hong Kong, Hong Kong, China}


\author{Wenxian Zhang}
\email[Corresponding email: ]{wxzhang@whu.edu.cn}
\affiliation{School of Physics and Technology, Wuhan University, Wuhan, Hubei 430072, China}

\date{\today}
\begin{abstract}
We theoretically investigate four types of dynamical instability, in particular the periodic and oscillatory type $I_O$, in an anti-ferromagnetic spin-1 Bose-Einstein condensate in a nonzero magnetic field, by employing the coupled-mode theory and numerical method. This is in sharp contrast to the dynamical stability of the same system in zero field. Remarkably, a pattern transition from a periodic dynamical instability $I_O$ to a uniform one $III_O$ occurs at a critical magnetic field. All the four types of dynamical instability and the pattern transition are ready to be detected in $^{23}$Na condensates within the availability of the current experimental techniques.
\end{abstract}


\maketitle
\section{introduction}
\begin{figure}
  \includegraphics[width=3.2in]{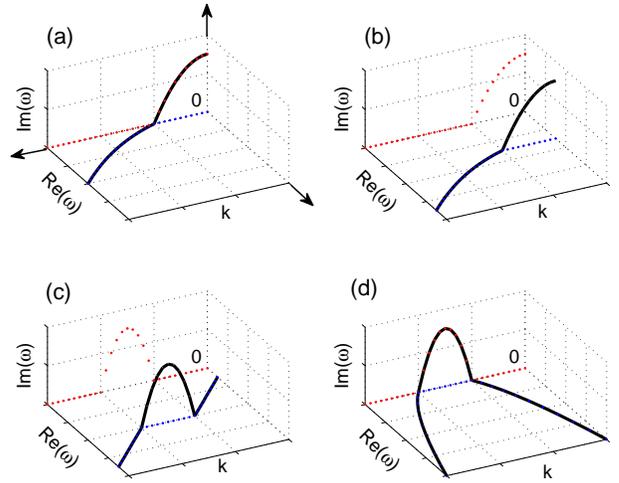}
  \caption{\label{fig:unstable}(Color online) Schematic diagram for four types of dynamical instability. The blue and the red dotted lines are the real and the imaginary part of the black solid line $\omega$, respectively. (a) Type $III_s$ dynamical instability appears with the peak locating at zero wave vector $k_p$ and zero $\re(\omega_p)$, (b) type $III_o$ at zero momentum $k_p$ and nonzero $\re(\omega_p)$, (c) type $I_o$ at nonzero momentum $k_p$ and nonzero $\re(\omega_p)$, (d) type $I_s$ at nonzero momentum $k_p$ and zero $\re(\omega_p)$.}
\end{figure}

Dynamical instabilities (DIs) exist in a wide variety of classical and quantum systems, such as solid state systems~\cite{PhysRevLett.63.1954}, liquid crystal~\cite{Dubois78}, nonlinear optics~\cite{Moloneybook}, chemistry~\cite{Mikhailov84, PhysRevLett.64.2953}, and biological systems~\cite{Tyson88}. The classical systems are well described by the linear response theory~\cite{RevModPhys.65.851, PhysRevB.54.16470}. According to this theory, four types dynamical instabilities are ideally distinguished, in terms of the characteristic wave vector $k_p$ and/or the real part of the frequency ${\rm Re}(\omega_p)$:
\begin{itemize}
  \item Type $III_S$ ($k_p = 0, {\rm Re}(\omega_p)=0$) DIs are uniform in space and stationary in time, as shown in Fig.~\ref{fig:unstable}(a). This DI type is usually considered trivial.
  \item Type $III_O$ ($k_p = 0, {\rm Re}(\omega_p)\neq 0$) DIs are uniform in space and oscillatory in time, as shown in Fig.~\ref{fig:unstable}(b). These type systems do not exhibit any spatial structure during its evolution.
  \item Type $I_O$ ($k_p \neq 0, {\rm Re}(\omega_p)\neq 0$) DIs are periodic in space and oscillatory in time, as shown in Fig.~\ref{fig:unstable}(c). This type systems spontaneously form a spatial pattern during its evolution, even starting from a uniform initial state.
  \item Type $I_S$ ($k_p \neq 0, {\rm Re}(\omega_p)= 0$) DIs are periodic in space and stationary in time, as shown in Fig.~\ref{fig:unstable}(d). This type systems may exhibit many complex spatial patterns which are actually decomposed into simple roll states with different wave vector $k$.
\end{itemize}

Quantum systems may exhibit various types DIs, in particular, in Bose-Einstein condensates (BECs), which are generally described by the Gross-Pitaevskii equation or nonlinear Schr\"odinger equation within the mean field theory\cite{RevModPhys.71.463, Kawaguchi2012253}. The fine tunability and the wonderful controllability in BEC experiments provide an excellent testbed for many theoretical predictions~\cite{RevModPhys.85.1191, PhysRevLett.114.225302, 2015PhRvL.114y5301L, PhysRevA.91.033635, 2015arXiv151206914A, Chapman15b, PhysRevA.92.023615, Eto14}. In fact, the DIs exist not only in scalar (single-component) BECs, such as solitons~\cite{PhysRevLett.83.5198} and vortices~\cite{PhysRevLett.83.2498}, but also in spinor BECs, including the two-component and three-component spinor BECs, where vector solitons~\cite{PhysRevA.79.033630}, Skyrmion vortices~\cite{Ruostekoski:2001fc, Khawaja01}, and spin domains/textures are explored~\cite{miesner1999observation, Sadler06, Vengalattore08, Vengalattore10,  2015arXiv151101624E, 2015arXiv151201331E, Tasgal15}. These DI types belong to either $III_O$ or $I_O$. The more complicated DI type $I_O$ was theoretically investigated only in two-component/two-mode BECs~\cite{bernier2014unstable} and experimentally in five-component spin-2 $^{87}$Rb condensates~\cite{kronjager2010spontaneous, Klempt09}. However, for the experimentally easily available spin-1 BECs, few efforts had been made in this direction.

In this paper, we systematically investigate all the four types DIs in an antiferromagnetic spin-1 BEC in a magnetic field. Although the antiferromagnetic spin-1 BEC is dynamically stable at zero magnetic field, the introduction of a nonzero magnetic field causes the system to exhibit all four types DIs. Unlike the unavoided crossing theory in the coupled two-component BECs, the emergence of the $I_O$ type DI requires the simultaneous coupling of the three modes in the spin-1 BEC. Moreover, an amazing pattern transition from type $I_O$ to type $III_O$ occurs as we increase the magnetic field. Our numerical calculations indicate that these types DIs are readily observed in $^{23}$Na condensates under current experimental conditions. The theory and methods open a door to the understanding of the spin texture observed experimentally in spin-1 condensates in nonzero magnetic fields with/without magnetic dipolar interaction~\cite{Sadler06, Vengalattore08, Vengalattore10}.

The paper is organized as follows. In Sec. II, we review the theoretical description of system and Number-conserving Bogoliubov theory .In Sec. III, we analytically design and numerically confirm magnetic field induced dynamical instabilities and Pattern transition.  In Sec. IV we present experimental consideration,and then give a conclusion.

\section{System description and Number-conserving Bogoliubov theory}
We consider a uniform anti-ferromagnetic spin-1 condensate in an external magnetic field ${\bf B}$ along the z axis. The Hamiltonian of the system is\cite{ho1998spinor,ohmi1998bose,law1998quantum, Zhang03}
\begin{eqnarray}\label{eq:ha}
  \hat H &=& \int\mathrm{d}\vec{r}\left[\hat\psi_{i}^{\dag}(\frac{-\hbar^{2}}{2M}\nabla^{2}+E_{i})\hat\psi_{i}+ \frac{c_{0}}{2}\hat\psi_{i}^{\dag}\hat\psi_{j}^{\dag}\hat\psi_{j}\hat\psi_{i} \right.\nonumber \\
   && \left.+\frac{c_{2}}{2}\hat\psi_{k}^{\dag}\hat\psi_{i}^{\dag}(F_{\gamma})_{ij}(F_{\gamma})_{kl}\hat\psi_{j}\hat\psi_{l}\right],
\end{eqnarray}
where $i,j,k,l \in \{\pm,0\}$ with $\pm,0$ denoting the magnetic quantum numbers $\pm 1,0$, respectively. Repeated indices are summed. $\psi_{i}(\psi_{i}^{\dag})$ is the field operator which annihilates (creates) an atom in the $i$th hyperfine state $|i\rangle \equiv |F=1,m_F=i\rangle$. $M$ is the mass of the atom. Interaction terms with coefficients $c_{0}$ and  $c_{2}$ describe elastic collisions of spin-1 atoms, namely, $c_{0}=4\pi \hbar^{2}(a_{0}+2a_{2})/3M$ and $c_{2}=4\pi \hbar^{2}(a_{2}-a_{0})/3M$ with $a_{0}$ and $a_{2}$ being the $s$-wave scattering lengths in singlet and quintuplet channels. The spin exchange interaction is anti-ferromagnetic (ferromagnetic) if $c_{2}>0 $ ($<0$). We focus on the anti-ferromagnetic spin interaction in this work. $F_{\gamma=x,y,z}$ are spin-1 matrices. $E_{i}$ denotes the Zeeman shift of an alkali atom in the state $|i\rangle$ (the Breit-Rabi formula)~\cite{vanier1989quantum,zhang2005coherent},
$E_{\pm} = -(E_{HFS}/{8})\mp g_{I}\mu_{I}B-(E_{HFS}/{2})\sqrt{1\pm \alpha+\alpha^2} $
and
$E_{0} = -(E_{HFS}/{8})-(E_{HFS}/{2}) \sqrt{1+\alpha^2}$,
where $E_{HFS}$ is the hyperfine splitting and $\alpha =(g_{I}\mu_{I}B + g_{J}\mu_{B}B)/E_{HFS}$. Here $g_{J}$ is the Land\'e $g$-factor for a valence electron with the total angular momentum $J=1/2$ and $g_{I}$ is the Land\'e $g$-factor for an atom with nuclear spin $I=3/2$, such as $^{87}$Rb and $^{23}$Na atoms. $\mu_{B}$ ($\mu_{I}$) is the Bohr (nuclear) magneton. For convenience we introduce the linear and the quadratic Zeeman term $\eta=(E_{-}-E_{+})/2$ and $\delta=(E_{+}+E_{-}-2E_{0})/2$, respectively.

To explore the DIs of the anti-ferromagnetic condensate in a nonzero magnetic field, we start from a stationary state, which is a solution to the following coupled equations~\cite{zhang2005coherent}, $
{c_2 n}\rho_{0}\sqrt{(1-\rho_{0})^{2}-{\sf m}^{2}}\sin \theta = 0$
and
$-{\delta}+{c_2n}(1-2\rho_{0})
 +{c_2n}\left[(1-\rho_{0})(1-2\rho_{0})-{\sf m}^{2}\right]/{\sqrt{(1-\rho_{0})^{2}-{\sf m}^2}}\cos \theta = 0$,
where $n$ is the total density of the condensate. The fractional population of the $i$th component is $\rho_{i} = n_i/n$ with $n_i$ the $i$th component density and $\sum_i n_i = n$, ${\sf m} = \rho_{+}-\rho_{-}$ is condensate magnetization, and $\theta = \theta_{+}+\theta_{-}-2\theta_{0}$ is the relative phase with $\theta_i$ being the phase of the $i$th component. The condensate spin wave function is $\xi_{i} = \sqrt{\rho_{i}}e^{-i\theta_{i}}$ and the energy of the condensate is~\cite{yi2002single,law1998quantum,pu1999spin}
$\varepsilon=c_2n\rho_{0}[(1-\rho_{0})+\sqrt{(1-\rho_{0})^{2} - {\sf m}^{2}}\cos\theta]+\delta(1-\rho_{0}) $. We limit ourselves to the stationary states with ${\sf m}=0$ and $\theta=0$, i.e., ${\bf \xi}=(\xi_{1}\  \xi_{0}\  \xi_{-1})^{\mathrm{T}}$ with
\begin{eqnarray}
  \xi_{\pm} &=& \sqrt{\frac{1}{4}+\frac{\delta}{8c_2n}}, \quad
  \xi_{0} = \sqrt{\frac{1}{2}-\frac{\delta}{4c_2n}} .
\end{eqnarray}
We focus on the three-component condensates, thus $\delta / c_2n \in (-2,2)$. In fact, the parameter range we explore in this work is $\delta/c_2n \ll 1$. We note that such a stationary state lies at a maximum point on the energy surface and is easily prepared by rotating a full polarized condensate to the $+x$ direction with a rf pulse in experiments~\cite{PhysRevLett.78.582}.

For a uniform spinor condensate, it is more convenient to work in momentum space by expanding the field operators in terms of plane waves as
$\hat\psi_{m} = \Omega^{-{1}/{2}} \sum_{\bf{k}}^{}e^{i\bf{k}\cdot\bf{r}}\hat a_{{\bf{k}},m}$
where $\Omega$ is the volume of the condensate and $\hat a_{{\bf k},m}$ ($\hat a_{{\bf k},m}^{\dag}$) is the annihilation (creation) operator of an atom with wave number $\bf k$ and magnetic number $m \in \{\pm 1,0\}$. Following the same way as in Ref.~\cite{ueda2000many, murata2007broken}, the original Hamiltonian Eq.~(\ref{eq:ha}) is rewritten as
$
  \hat H = \sum\limits_{{\bf k},m}^{}(\epsilon_{\bf k}-\eta m+\delta m^{2})\hat a_{{\bf k},m}^{\dag}\hat a_{{\bf k},m} +(c_{0}/2\Omega)\sum\limits_{{\bf k}}^{}:\hat\rho_{\bf k}^{\dag}\hat\rho_{\bf k}:+(c_{2}/2\Omega)\sum\limits_{{\bf k}}^{}:\hat f_{\bf k}^{\dag}\hat f_{\bf k}:,
$
where $\epsilon_{\bf k}=\hbar^2{\bf k}^2/(2M)$ is the kinetic energy of the collect excitation mode with wave vector
$\bf k$, $\hat\rho_{\bf k}\equiv\sum_{{\bf q},m}^{}\hat a_{{\bf q},m}^{\dag}\hat a_{{\bf q+k},m}$ and $\hat f_{\bf k}\equiv\sum_{{\bf q},m,n}^{}{\bf f} _{m,n}\hat a_{{\bf q},m}^{\dag}\hat a_{{\bf q+k},n}$ with ${\bf f}=(F_{x},F_{y},F_{z})$ being the spin-1 matrices in vector notation. The symbol $::$ denotes the normal ordering of operators.

The dispersion relations of the system can be directly calculated with the number-conserving Bogoliubov theory, which preserves the atom number $N$ without introducing the chemical potential as Lagrange multiplier. According to this theory, by substituting $\hat a_{{\bf 0},m}$ with
$\xi_{m}(N-\sum_{{\bf k\neq 0},m}^{}\hat a_{{\bf k},m}^{\dag}\hat a_{{\bf k},m})^{1/2}$
and keeping terms up to the second order in $\hat a_{{\bf k\neq 0},m}$ ($\hat a_{{\bf k\neq 0},m}^{\dag}$), the effective Bogoliubov Hamiltonian becomes~\cite{murata2007broken}
\begin{widetext}
\begin{eqnarray}\label{eq:heff}
   \hat H_e &=& \sum\limits_{{\bf k\neq 0}}^{} \sum\limits_{ m=-1}^{1}[\epsilon_{\bf k}-\eta m+\delta m^{2}+\eta\langle F_{z} \rangle-\delta\langle F_{z}^{2} \rangle
   - c_{2}n(1-|\xi_{0}^{2}-2\xi_{1}\xi_{-1}|^{2})]\hat a_{{\bf k},m}^{\dag}\hat a_{{\bf k},m} + c_{2}n\langle {\bf f} \rangle\cdot\sum\limits_{{\bf k\neq 0}}^{}\sum\limits_{m,n}^{}{\bf f}_{m,n}\hat a_{{\bf k},m}^{\dag}\hat a_{{\bf k},m} \nonumber \\
   &+& \frac{c_{0}n}{2}\sum\limits_{{\bf k\neq 0}}^{}(2\hat D_{\bf k}^{\dag}\hat D _{\bf k}+\hat D_{\bf k}\hat D_{-\bf k}+\hat D_{\bf k}^{\dag}\hat D_{-\bf k}^{\dag}) + \frac{c_{2}n}{2}\sum\limits_{{\bf k\neq 0}}^{}(2\hat{\bf \digamma}_k^{\dag} \cdot \hat{\bf \digamma}_k+\hat{\bf \digamma}_k \cdot \hat{\bf \digamma}_{-k}+\hat{\bf \digamma}_ k^{\dag} \cdot \hat{\bf \digamma}_{-k}^{\dag}),
\end{eqnarray}
where $n=N/\Omega$, $D_{\bf k}=\sum_{m}^{}\xi_{m}^{\ast}\hat a_{{\bf k},m}$, $\hat{\bf \digamma}_k=\sum_{m,n}^{}{\bf f}_{m,n}\xi_{m}^{\ast}\hat a_{{\bf k},n}$, $\langle F_{z}\rangle$ ($\langle F_{z}^{2}\rangle$) is the average of $F_{z}(F_{z}^{2})$ over the stationary state $\bf{\xi}=(\xi_{1}\  \xi_{0}\  \xi_{-1})^{\mathrm{T}}$ and the constant term is ignored.

The square of the Bogoliubov excitation spectrum $\omega_{{\bf k},\sigma}^2$ is straightforwardly calculated as the eigenvalues of the non-hermitian matrix
\begin{equation}\label{}
G = (M+N)(M-N)
\end{equation}
where
\begin{eqnarray}
   M &=&\begin{pmatrix}
          A_{1} & c_{2}n(\xi_{1}\xi_{0}^{*}+2\xi_{0}\xi_{-1}^{*})+c_{0}n\xi_{1}\xi_{0}^{*} & c_{0}n\xi_{1}\xi_{-1}^{*}-c_{2}n\xi_{1}\xi_{-1}^{*} \\
          c_{2}n(\xi_{0}\xi_{1}^{*}+2\xi_{-1}\xi_{0}^{*})+c_{0}n\xi_{0}\xi_{1}^{*} & A_{0} & c_{2}n(2\xi_{1}\xi_{0}^{*}+\xi_{0}\xi_{-1}^{*})+c_{0}n\xi_{0}\xi_{-1}^{*} \\
          c_{0}n\xi_{-1}\xi_{1}^{*}-c_{2}n\xi_{-1}\xi_{1}^{*} & c_{2}n(2\xi_{0}\xi_{1}^{*}+\xi_{-1}\xi_{0}^{*})+c_{0}n\xi_{-1}\xi_{0}^{*}  & A_{-1}
        \end{pmatrix} \nonumber\\
   & &  \nonumber\\
   N &=&n\begin{pmatrix}
          (c_{0}+c_{2})\xi_{1}^{2} & (c_{0}+c_{2})\xi_{1}\xi_{0} & (c_{0}-c_{2})\xi_{1}\xi_{-1} +c_{2}\xi_{0}^{2} \\
          (c_{0}+c_{2})\xi_{1}\xi_{0}  & c_{2}\xi_{0}^{2}+2c_{2}\xi_{1}\xi_{-1} & (c_{0}+c_{2})\xi_{0}\xi_{3} \\
          (c_{0}-c_{2})\xi_{1}\xi_{-1} +c_{2}\xi_{0}^{2} & (c_{0}+c_{2})\xi_{0}\xi_{3} & (c_{0}+c_{2})\xi_{1}^{2}
        \end{pmatrix} \nonumber
 \end{eqnarray}
with
$A_{m}=\epsilon_{k}-\eta m+\delta m^{2}+\eta\langle F_{z}\rangle
        -\delta\langle F_{z}^{2}\rangle+c_{2}n
        (|\xi_{m}|^2-2|\xi_{-m}|^2+|\xi_{0}^{2}-2\xi_{1}\xi_{-1}|^{2}) +c_{0}n|\xi_{m}|^{2}$.

\begin{figure}
\centering
\includegraphics[width=6in]{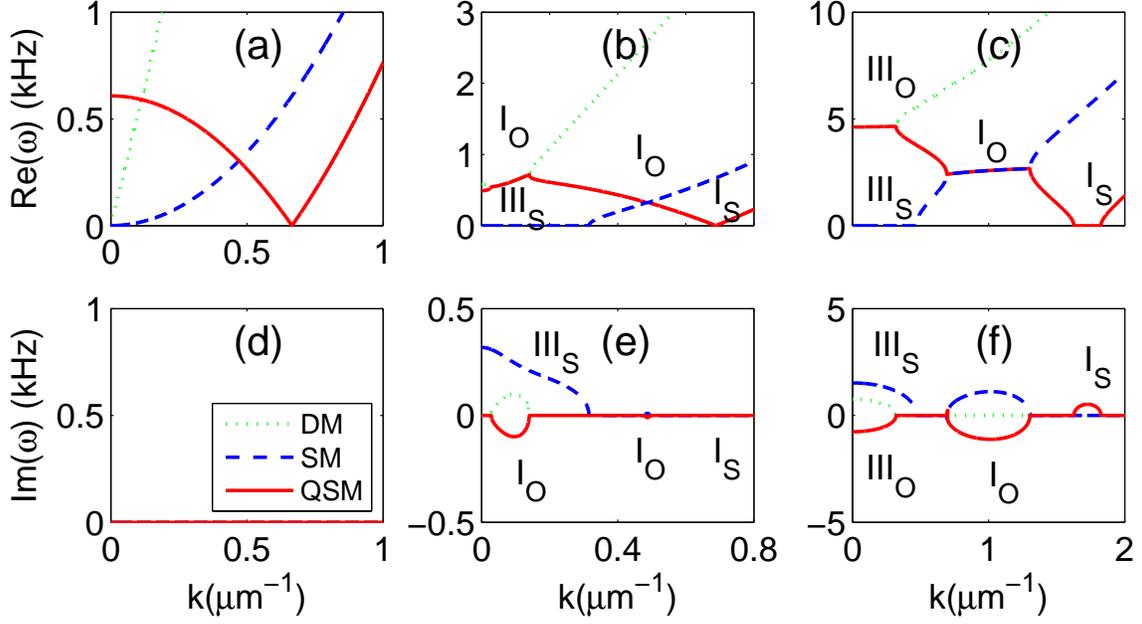}
\caption{\label{fig:mode}(Color online) Dispersion relations in an anti-ferromagnetic spin-1 BEC. Only the positive branches are plotted. Real (a) and imaginary (d) parts of the collective modes' frequency at $B=0$. Real (b) and imaginary (e) parts of the collective modes' frequency at $B = 0.04$ mG. Real (b) and imaginary (e) parts of the collective modes' frequency at $B = 1$ mG. DM, SM, and QSM are denoted with green dotted lines, blue dashed lines, and red solid lines, respectively. The condensate density is $n=10^{14}$ cm$^{-3}$. Four types of DIs all appear in a large nonzero magnetic field.}
\end{figure}
\end{widetext}

\section{Magnetic field induced dynamical instabilities and Pattern transition}
\subsection{Magnetic field induced dynamical instabilities}
\begin{figure}
\centering
\includegraphics[width=1\linewidth]{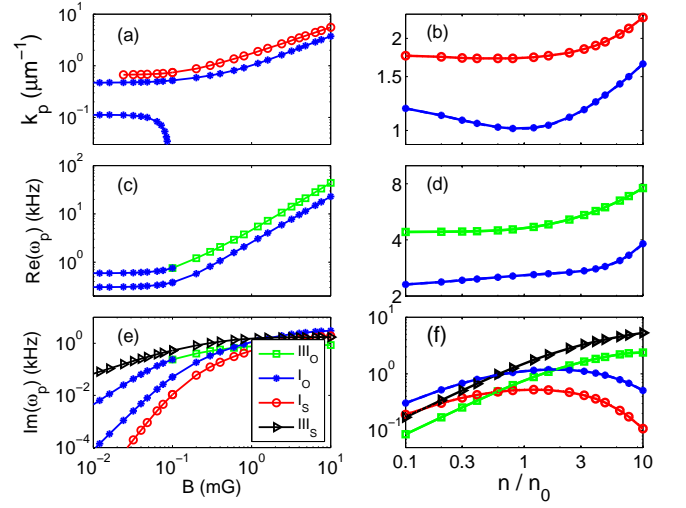}
\caption{\label{fig:range}(Color online) Dependence of the peak positions (top row), the real part (middle row) and the imaginary part (bottom row) of the peak values of four types unstable modes on the magnetic field (left column) and on the condensate density (right column) in an antiferromagnetic spin-1 $^{23}$Na condensate. The density is $n=n_0$ with $n_{0} = 10^{14} \; {\rm cm}^{-3}$ for the left column and the magnetic field is $B = 1$ mG for the right column. The magnetic fields and the densities are available in current $^{23}$Na condensate experiments.}
\end{figure}

We numerically calculate the dispersion relations for a uniform $^{23}$Na condensate in zero and nonezero magnetic fields and present the results in Fig.~\ref{fig:range}. At $B=0$, as shown in Fig.~\ref{fig:mode}(a) and (d), the imaginary part of the frequency is zero for all three modes, the density mode (DM), the spin mode (SM), and the quadrupolar spin mode (QSM). These zero imaginary parts indicate that the stationary state is dynamically stable, consistent with previous results~\cite{zhang2005dynamical}. Such an agreement indicates that the number-conserving Bogoliubov theory essentially produces the same collective excitation spectrum as the standard one but is more convenient without introducing additional Lagrange multiplier~\cite{murata2007broken}. There are four interesting cross points as the wave vector $k$ increases in Fig.~\ref{fig:mode}(a): (i) the SM and the DM crosses at the origin; (ii) the QSM crosses the DM; (iii) the QSM crosses the SM; (iv) the QSM touches zero frequency. All the last three cross points lie at nonzero wave vectors. Dynamical instability might occur near these four cross points if a perturbation is purposely introduced~\cite{bernier2014unstable}, as we numerically prove in the following.

Typical types DIs are presented in Fig.~\ref{fig:mode}(b) and (e), and (c) and (f), for $B=0.04$ mG and $1$ mG, respectively. From Fig.~\ref{fig:mode}(b) and (e), three DI types are observed at $B=0.04$ mG: $III_S$ at $k=0$, two $I_O$'s and $I_S$ as $k$ increases. From Fig.~\ref{fig:mode}(c) and (f), there are four DI types, $III_S$ and $III_O$ at $k=0$, $I_O$ and $I_S$ as $k$ increases.

By comparing Fig.~\ref{fig:mode}(d) and (e), different types DIs indeed occur by introducing a small but nonzero magnetic field and the peak positions of the wave vector of the DI correspond, respectively, to the cross points in Fig.~\ref{fig:mode}(a). As shall be shown, the emergence of these DIs is clearly explained by the following perturbation theory, where the magnetic field effects are treated as perturbation.

At $B=0$, we define $G=G_0$. It is easy to find the eigenvalues of $G_0$ and the transformation matrix $S$,
\begin{eqnarray}
  S^{-1}G_0S &=& \begin{pmatrix}
          \omega_S^2 & 0 & 0 \\
          0& \omega_Q^2 & 0 \\
          0 & 0 & \omega_D^2
        \end{pmatrix} \nonumber
\end{eqnarray}
with $\omega_S = \epsilon_{k}$, $\omega_Q = \epsilon_{k}-2c_{2}n$, and $\omega_D = \sqrt{[\epsilon_{k}+2n(c_{0}+c_{2})]\epsilon_{k}}$,
where
\begin{eqnarray}
  S &=& \frac 1 2\begin{pmatrix}
          -\sqrt 2 & 1 & 1 \\
          0 & -\sqrt{2} & \sqrt{2} \\
          \sqrt 2 & 1 & 1
        \end{pmatrix}. \nonumber
\end{eqnarray}
These eigenmodes are SM, QSM, and DM.

At a small magnetic field, the stationary state is approximately the same as $B=0$ since the quadratic Zeeman effect $\delta/(c_2n)$ is negligible. The perturbation term introduced by the magnetic field $V=S^{-1}(G-G_0)S$ is, expressed in the eigenbasis of $G_0$,
\begin{eqnarray}
V &=& \eta \begin{pmatrix}
               \eta &\sqrt 2(\epsilon_{k} - c_2n) & \sqrt 2\epsilon_{k} \\
               \sqrt 2(\epsilon_{k} - c_2n) & \eta/2 & \eta/2 \\
               \sqrt 2[(c_0 + c_2)n + \epsilon_{k}] & \eta/2 & \eta/2
             \end{pmatrix}. \nonumber
\end{eqnarray}
Clearly, the magnetic field terms couple all the three modes and cause various types of DIs, including the trivial $III_S$ pattern, the common $I_S$~\cite{matuszewski2010rotonlike}, and the long-sought $I_O$'s~\cite{zhang2005dynamical,bernier2014unstable}.

\begin{figure}
\centering
\includegraphics[width=1\linewidth]{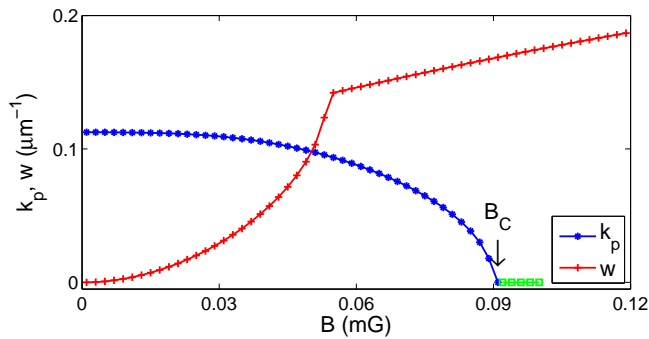}
\caption{\label{fig:crit}(Color online) Pattern transition. The DI type changes from $I_O$ to $III_O$ as $B>B_C$, indicated by the peak position lying at $k_p=0$. The width of the DI region changes drastically from a quadratic function of $B$ to a linear one before $B_C$.}
\end{figure}

At a relatively large magnetic field, by comparing Fig.~\ref{fig:mode}(e) and (f), we find that the DIs become more significant with larger imaginary parts. To systematically investigate the magnetic field effect on these DIs, at each magnetic field, we extract the peak position of the imaginary part of the four types DIs, $k_p$, and the corresponding real and imaginary part of the frequency, $\omega_p$. These results are presented in the left column of Fig.~\ref{fig:range}, from which we find that the peak positions, the real part and the imaginary part of the peak values all increase monotonically with the magnetic field, except the lower-$k_p$ $I_O$ whose peak position decreases down to zero (see also Fig.~\ref{fig:crit}).

The density dependence of the peak properties is presented in the right column of Fig.~\ref{fig:range}. The peak positions of $I_O$ and $I_S$ decrease a little and then increase as the density increases. Importantly, the lowest $k_p$ is about $1\; \mu m^{-1}$ at the density of $n=10^{14}$ cm$^{-3}$, which implies that a spatial pattern with a characteristic scale of $2\pi / k_p \sim 6 \; \mu m$ appears. Such a spatial pattern is experimentally detectable and the atom density is readily available with current experimental techniques~\cite{stamper2014seeing}. From Fig.~\ref{fig:range}(d), we find that the real parts of the peak value for $III_O$ and $I_O$ increase with the density. Interestingly, the imaginary parts for the $I_O$ and $I_S$ increase at low densities, but decrease at high densities, as shown in Fig.~\ref{fig:range}(f).

\subsection{Pattern transition}
The unusual decreasing of peak position of the lower-$k_p$ $I_O$ to zero in Fig.~\ref{fig:range}(a) and Fig.~\ref{fig:crit} actually manifests a remarkable pattern transition to $III_O$. Such a pattern transition has not been revealed before, to our best knowledge. As shown in Fig.~\ref{fig:crit}, the width of the $I_O$ region {\sf w} increases rapidly in a rough quadratic form but the peak position is almost fixed if the magnetic field is smaller than 0.06 mG. Once the left end of the $I_O$ region touches the origin $k=0$, the width changes only linearly with $B$ but the peak position decreases drastically down to zero. The DI type becomes $III_O$ since $k_p=0$, if $B>B_C$.

\section{Conclusion}
To experimentally observe the DIs and the pattern transition, on one hand, it is important to lower the magnetic field so that the characteristic length of the DIs is larger than the spatial resolution of about 1 $\mu$m for the detector. Within a magnetically shielded room, the magnetic field can reach as low as $10^{-2}$ mG with phase compensation technique~\cite{Eto13}, which is low enough for the DIs and the pattern transition. On the other hand, the characteristic time of the DIs [$1/Im(\omega_p)$] must be shorter than the spin-1 condensate life time which can reach as long as $10^2$ s~\cite{RQWang}. One has to balance the two factors in a practical $^{23}$Na condensate experiment.


We predict all the four types DIs in an anti-ferromagnetically interacting spin-1 condensate in a magnetic field with the number-conserving Bogoliubov theory. Remarkablely, the system exhibits a pattern transition from $I_O$ to $III_O$ once the magnetic field crosses the critical value $B_c$. Our theoretical predictions about the DIs and the pattern transition are readily to be verified in a $^{23}$Na spin-1 condensate under currently available experimental conditions.

\acknowledgments
W.Z. and D.W. thank L. You for inspiring discussions at the early stage of this work. This work is supported by the National Natural Science Foundation of China Grant No. 11574239, 11434011, 11547310, and 11275139, the National Basic Research Program of China Grant No. 2013CB922003, RGC Hong Kong (GRF CUHK403813), and the Fundamental Research Funds for the Central Universities.



\end{document}